\documentclass[12pt]{article}

\usepackage{amsmath}
\usepackage{amssymb}

\usepackage{graphicx}
\usepackage{graphicx}
\usepackage{cite}

\usepackage[usenames,dvipsnames]{xcolor}

\usepackage[labelfont=bf,labelsep=period,justification=raggedright]{caption}

\makeatletter
\renewcommand{\@biblabel}[1]{\quad#1.}
\makeatother

\date{}

\begin{document}

\begin{flushleft}
{\Large
\textbf{Effects of Mixing Interaction Types on Ecological Community Stability.}
}
\\
Samir Suweis$^{1,\ast}$, 
Jacopo Grilli$^{1,\ast}$, 
Amos Maritan$^{1}$
\\
\bf{1} Dipartimento di Fisica e Astronomia, Universit\`a di Padova, CNISM and INFN, via Marzolo 8, 35131 Padova, Italy
\\
$\ast$ \small{These two authors contributed equally- Emails: : suweis@pd.infn.it; jgrilli@pd.infn.it}
\end{flushleft}

\section*{Abstract}
In the last years, a remarkable theoretical effort has been made in order to understand stability and complexity in ecological communities.
The non-random structures of real ecological interaction networks has been recognized as one key ingredient contributing to the coexistence between high complexity and stability in real ecosystems. However most of the theoretical studies have considered communities with only one interaction type (either
antagonistic, competitive, or mutualistic). Recently it has been proposed a theoretical analysis on multiple interaction types in ecological
systems, concluding that: $a)$ Mixture of antagonistic
and mutualistic interactions stabilize the system with respect to the less realistic
case of only one interaction type; $b)$ Complexity, measured in terms of the
number of interactions and the number of species, increases stability in systems with different
types of interactions. By introducing new theoretical investigations and
analyzing 21 empirical data sets representing mutualistic plant-pollinator networks, we show that
that conclusions are incorrect. We will prove that the positive complexity-stability effect observed in systems with different kind of interactions is a mere consequence of a rescaling of the interaction strengths, and thus unrelated to the mixing of interaction types. 

\section*{Introduction}
The relationship between diversity and stability in ecosystems is one of the most debated issue by ecologists \cite{May1972,May1974,Pimm1984,McCann2000,Ives2007,Bastolla2009,Allesina2012}. Empirical evidences \cite{Pimm1984,Ives2007}  suggest a positive diversity - stability relationship, while theoretical studies challenge this point of view. The discussion started since publication of the pioneer work by Robert May \cite{May1972} which provides a quantitative relation of th community stability of randomly assembled ecosystems with its number of species ($S$), connectance ($C$) - i.e. the ratio between actual and potential interactions in the ecosystem - and the characteristic interaction strength.
Specifically, for random interactions drawn from a Gaussian distribution of zero mean and standard deviation $\sigma$, a randomly assembled system is stable (i.e. all real part of the eigenvalues of the stability matrix  are negative) if
\begin{equation}\label{may_rel}
\displaystyle
\sigma\sqrt{C S} < 1
 \ ,
\end{equation}

This result rises the paradox  that more complex an ecosystem is (i.e. large $C\cdot S$), less stable it is \cite{May1974}. This important result suggests that real networks must have some non-random, stabilizing structures that allow ecosystems to persist despite their complexity \cite{Fox2006,Bascompte2010}. The structure and the interactions types (e.g. mutualistic, antagonistic, etc..) have indeed a remarkable impact on the stability profiles of ecosystem community dynamics \cite{Bascompte2007,Thebault2010,Allesina2012}. In particular it has been found that predator-prey systems are more stable with respect to random assembled community, while mutualistic communities are not \cite{Allesina2008,Allesina2012}. However, these results do not solve the paradox: still for structured community ecosystem stability decreases for increasing $C$ or $S$ (and fixed interaction strength) \cite{Allesina2012}.

A possible solution to the paradox has been recently proposed~\cite{Mougi2012,Boyd2012}.
A theoretical analysis on multiple interaction types in ecological systems~\cite{Mougi2012} concluded that a mixture of antagonistic and mutualistic interactions is more likely to stabilize the system with respect to the less realistic case of interactions of a single type. Moreover it was found that complexity, measured in terms of the connectance and the number of species, increases stability in systems with different types of interactions, in contrast with May's result \cite{May1974}. 

In this work we show that the mixing of interaction types does not solve the paradox in the stability-diversity relationship. On the contrary, also for an ecosystem with hybrid interactions, stability decreases as  complexity increases. We will first briefly summarize  the theoretical framework and mathematical details of the community dynamics model used by Mougi and Kondoh \cite{Mougi2012}  and variants of it. We will then show that: $a)$ Dataset of real mutualistic interaction networks do no support a key assumption made in their framework; $b)$ Even from a theoretical point of view a mixture of interaction types does not have a stabilizing effect; $c)$ The positive complexity-stability effect recently proposed \cite{Mougi2012} is only a consequence of a rescaling of interaction strengths, which is unrelated to the mixing of interaction types.


\section*{Materials and Methods}
\subsection*{Community Dynamics Model}

The model considered by Mougi and Kondoh \cite{Mougi2012} is formulated through Lotcka-Volterra dynamics \cite{Hamilton1989}
\begin{equation}
\displaystyle
\frac{d n_i}{d t} = n_i \Bigl( r_i + \sum_{j=1}^S a_{ij} n_j  \Bigr) \ ,
\label{eq:LV-def}
\end{equation}
where $n_i$ is the abundance of species $i$, $S$ is the number of species, $r_i$ is the intrinsic rate of growth and
$a_{ij}$ is the interaction coefficient between species $i$ and species $j$ while $a_{ii}=z_i$ represents the self interaction (density dependent regulation) and it is uniformly distributed between $0$ and $1$ ($z_i\sim U_{[0,1]}$).
The matrix $a_{ij}$ contains all the information about the interactions between species. If species $i$ has a mutualistic interaction with species $j$ ($i \neq j$), the corresponding terms in the interaction matrix will have the form
\begin{equation}
\begin{split}
\displaystyle
a_{ij} & = f_M e_{ij} \frac{A_{ij}}{  \sum_{k \in m(i)} A_{ik}} \\
a_{ji} & = f_M e_{ji} \frac{A_{ji}}{  \sum_{k \in m(j)} A_{jk}} \ ,
\label{eq:mat-def_mut}
\end{split}
\end{equation}
while if species $i$ has an antagonistic interaction with $j$ (consider for instance $i$ predator and $j$ prey)
\begin{equation}
\begin{split}
\displaystyle
a_{ij} & = f_A g_{ij} \frac{A_{ij}}{  \sum_{k \in p(i) } A_{ik}} \\
a_{ji} & = - f_A  \frac{A_{ij}}{  \sum_{k \in p(i) } A_{ik}} = - \frac{a_{ij}}{g_{ij}} \ ,
\label{eq:mat-def_ant}
\end{split}
\end{equation}
where $m(i)$ is the set of the species having a mutualistic interaction with $i$, whereas $p(i)$ the species which are preys of $i$.
$A_{ij}$ measures the potential preference of an interaction between $i$ and $j$, $f_A$ and $f_M$ are the relative strengths of the antagonistic and mutualistic interaction, respectively,
while $e_{ij}$ and $g_{ij}$ quantifies the symmetry of the interactions. The model could also be generalized to a non-linear functional response (Holling type II) \cite{Thebault2010}.
Not all the species interact: only a fraction $C$ of matrix elements $A_{ij}$ is different from zero, of which a fraction $p_M$ is mutualistic, while a fraction $1-p_M$ is antagonistic.

For a given choice of $i)$ the structure of the matrix  $A_{ij}$ (random, cascade for antagonistic part and bipartite for mutualistic one)\cite{Allesina2012}, $ii)$ the parameters $f_A$ and $f_M$ and $iii)$ randomly drawn values of $e_{ij}$ and $g_{ij}$ uniformly distributed between 0 and 1,  Eqs.~(\ref{eq:mat-def_mut}) and~(\ref{eq:mat-def_ant})
lead to the interaction matrix $a$. By introducing a stationary point $\vec{n}*=(n_1,\dots,n_S)$, whose components are randomly drawn from a uniform distribution between $0$ and $1$, and linearizing Eq.~(\ref{eq:LV-def}) around $\vec{n}*$ one finally obtains the stability matrix $M$.
 If all eigenvalues of $M$ have negative real parts, the system is stable. Otherwise it is unstable.
Mougi and Kondoh~\cite{Mougi2012} have shown that starting from a mutualistic (antagonistic) interaction matrix and adding antagonistic (mutualistic) links, i.e.
decreasing (increasing) $p_M$, the stability increases in a non linear fashion. Moreover they also showed that for hybrid communities the stability increases as $S$ and/or $C$ increases.

The crucial assumption made in~\cite{Mougi2012}, is to impose a "constant interacting effort" for each species, i.e. if a species is generalist and thus can positively interact with several different species, then the average interaction strength must be smaller than the one of specialist species that has only few resources.  This fact should be translated mathematically by rescaling the interactions as
\begin{equation}
\displaystyle
a_{ij} \sim \frac{A_{ij}}{ \sum_{k  \in  m(i)\cup p(i)} A_{ik}} \sim
\frac{A_{ij}}{S C (p_M + \frac{1-p_M}{2}) \mathbb{E}(A)}
\label{eq:just-ass}
\end{equation}
where $m(i)\cup p(i)$ is the set of the resources species of the species $i$, i.e. both the mutualistic partners or preys of $i$. However, in the work Mougi and Kondoh the "constant interacting effort" is imposed in a more stringent way:  the interacting effort spent separately in mutualistic and in antagonistic interactions are fixed to a constant. They indeed define the rescaled interaction $a_{ij}$ between the species $i$ and the species $j$ as shown in Eqs~(\ref{eq:mat-def_mut}) and~(\ref{eq:mat-def_ant}).
 As a consequence of this stronger assumption, the average effort spent by a species in mutualistic or antagonistic interactions does not depend on $p_M$. The total effort spent by species $i$ to interact with its mutualistic partners is indeed $\sum_{j \in m(i)}  a_{ij} \sim f_M\mathbb{E}(e_{ij}A_{ij})/\mathbb{E}A_{ij}=f_M /2$, where $\mathbb{E}(\cdot)$ represents the expectation value, and we have used the law of large numbers and the fact $e_{ij}$ and $A_{ij}$ are independent random variables. 
In the same way the effort spent by the species $i$ as a predator is on average equal to $f_A /2$.
Therefore the authors of ref. \cite{Mougi2012} do not only assumes that, $(i)$ \textit{``interaction strengths decrease with increasing resource species, due to an allocation of interacting effort'', but also that, $(ii)$ the total interaction strengths spent in mutualistic (antagonistic) interactions does not depend on the number of mutualistic (antagonistic) interactions. $(i)$ and $(ii)$ are encapsulated by the denominators in Eqs.~(\ref{eq:mat-def_mut}) and~(\ref{eq:mat-def_ant}).} In other words they are assuming that the mutualistic interacting effort spent in a ecosystem where $1\%$ or $99\%$  of the links are mutualistic, is the same. 

\subsection*{Data Analysis}

\begin{figure*}[h!]
\centering
  \includegraphics[width=0.9\textwidth]{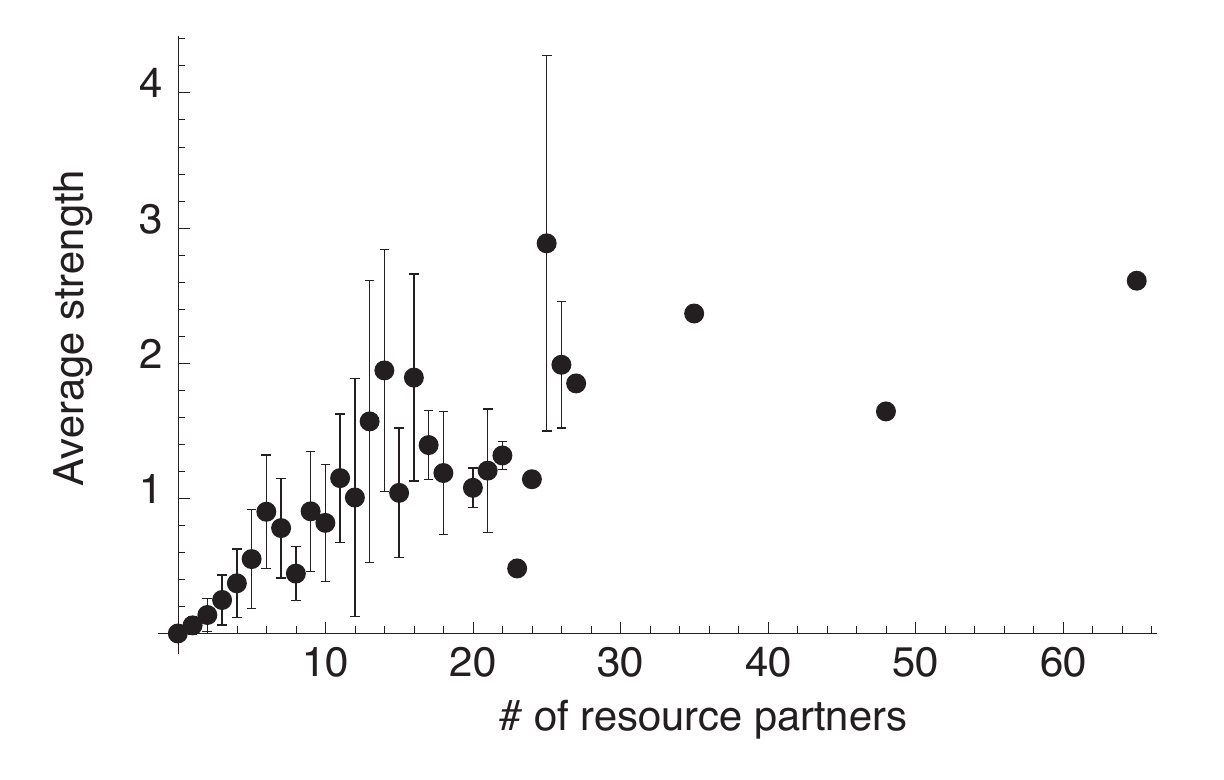}
  \caption{The interaction effort of each species is correlated to the number of interactions.
	The figure shows relation between the average interaction strength ($\bar{s}$) and the
	number of its resource partners ($d_i$). We average over the normalized strength of the species sharing the
	same value of $k$ (black points) among the databases. The error bars are calculated as the standard deviation of the average ($\sigma_s$). Points without error bar are single data points and thus they are less reliable.}
\label{fig:data}
\end{figure*}

Although at least assumption $(i)$ might seems plausible and  biologically justifiable, we show that this, and thus assumption $(ii)$, is not supported by observations in empirical mutualistic ecosystems (see Figure~\ref{fig:data}). 

We have analyzed a dataset consisting of $21$ empirical mutualistic networks~\cite{Jordano2007}.
All information on each dataset $b$ is incorporated in the weighted adjacency matrix $W^b$, whose elements $w^b_{ij}$ gives the interaction strength between species $i$ and $j$.  Therefore the normalized total interaction strength of species $i$ in database $b$ is defined as $s^b_i=\sum_{k \in m(i)} w^b_{ik}/\max_{kl}(w^b_{kl})$, while the number of interacting partners  is $d^b_i=\sum_{k \in m(i)} \Theta(w^b_{ik})$ ($\Theta(x)=1$ if $x=0$ and 0 otherwise). From these information we can analyze the empirical relations between $s$ and $d$, i.e. the relationships between the positive number of resources of a species and its interaction strength. In order to minimize the effect of fluctuations due to intrinsic stochastic variability in ecosystem dynamics and errors in interaction strength detection, we have averaged this relation across all databases, i.e.
\begin{equation}
\bar{s}(d)=\frac{\sum_b \sum_{i=1}^{S(b)} s^b_i \delta(d^b_i - d) }{\sum_b \sum_{i=1}^{S(b)} \delta(d^b_i - d)} \ ,
\end{equation}
where $S(b)$ is the number of species in the dataset $b$.

A constant interacting effort hypothesis would suggest that $\bar{s}$ and $d$ are anti-correlated: $\bar{s}\propto 1/d$. On the contrary we find that the average interaction strength and the number of interactions are strongly positive correlated, giving a falsification of assumptions $(i)$ and $(ii)$. We stress that, as also Mougi and Kondoh noted~\cite{Mougi2012}, the relaxation of assumption $(ii)$ is crucial for the validity of all their conclusions. The stabilizing effects due to the mixing of mutualistic and antagonistic interactions types vanish if the constant interaction effort hypothesis does not hold separately for the two kinds of interactions.

\section*{Discussion and Results}

We have shown that data do not support the assumption that interaction strengths decreases with increasing resources species. We now show that even from a pure theoretical ground it is neither true that multiple interaction types stabilize ecosystems nor that increasing complexity increases stability.

\subsection*{Stability Criteria for Hybrid Communities }

We first analyze the impact of mixing mutualistic  and predator-prey interactions type on the ecosystem stability by extending to hybrid communities the analysis recently proposed by Allesina and Tang \cite{Allesina2012}. In this subsection we thus go beyond the particular model dynamics presented above, and we use a more general approach based on .

Consider the $S\times S$ stability matrix $M$ for a given community dynamics $\dot{n}_i=f(\vec{n})$ with $i=1,...S$.  $M_{ij}$ describes the effect of the interaction between species $i$ and $j$ in the proximity of a feasible stationary point $\vec{n}^*$ of the underling dynamics $f$: $M_{ij}\equiv\frac{\partial f_i(\vec{n}_i) }{\partial n_j}_{|_{\vec{n}^*}}$. $M$ for hybrid predator-prey and mutualistic interactions networks can be built in the following way \cite{Allesina2012}.  We first pick at random a pair $i-j$ of species and we draw a random value $p$ from a uniform distribution between zero and one ($U_{[0,1]})$. If $p< C$, the these two species interact, otherwise they do not interact ($M_{ij}=M_{ji}=0$). If $i-j$ are interacting species, then with probability 1-$p_M$ species $i$ preys species $j$, otherwise they are mutualistic partners.
In the former case we set $M_{ij} \sim | \mathcal{N}(0,\sigma^2)|$ and $M_{ji}\sim -|\mathcal{N}(0,\sigma^2)|$, otherwise $M_{ij} \sim | \mathcal{N}(0,\sigma^2)|$ and $M_{ji}\sim |\mathcal{N}(0,\sigma^2)|$.  $\mathcal{N}(\mu,\sigma^2)$ is the normal distribution with mean $\mu$ and variance $\sigma^2$, and determines the intensity of the interactions among species (with the notation $|\mathcal{N}(0,\sigma^2)|$ we mean that a random number is taken from $\mathcal{N}(0,\sigma^2)$ and its modulus $|\cdot |$ is taken).  

\begin{figure*}[h!]
\centering
  \includegraphics[width=0.8\textwidth]{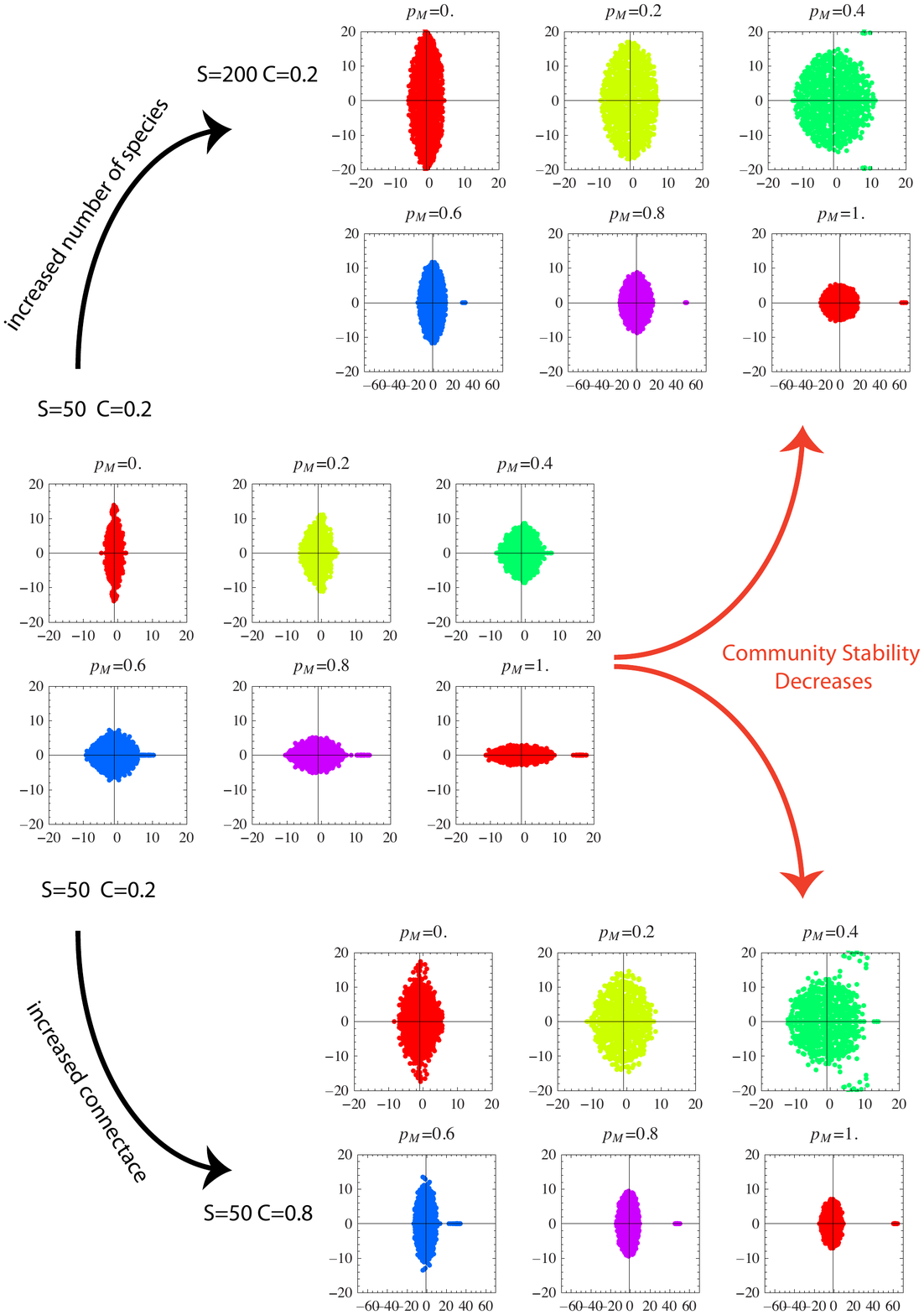}
  \caption{Stability profile of the stability matrix $M$ in the model independent framework presented in \cite{Allesina2012}. The linearized matrix $M$ are obtained as
	done in \cite{Allesina2012}.  The profile of eigenvalues distribution in the complex plane, shows that increasing complexity
	decreases stability, i.e. the maximum real part of the eigenvalues of $M$ increases for increasing $S$ and/or $C$. Moreover the stability is not increased by the mixing of interaction types, but as expected \cite{Allesina2012} adding mutualistic links decreases stability.}
\label{fig:comp-stab}
\end{figure*}

Following this simple algorithm one can build cascade predator-prey community matrices with a desired fraction $p_M$ of mutualistic interactions and then analyze the corresponding eigenvalues to study the stability for several levels of diversity of interaction type in the ecological community. Figure~\ref{fig:comp-stab} shows that also for hybrid community the stability of the linearized dynamics described by $M$ decreases as $S$ and/or $C$ increases, independently of $p_M$. Moreover, as expected from previous results \cite{Allesina2012}, mutualistic interactions destabilize the community. 

\begin{figure*}[h!]
\centering
  \includegraphics[width=0.9\textwidth]{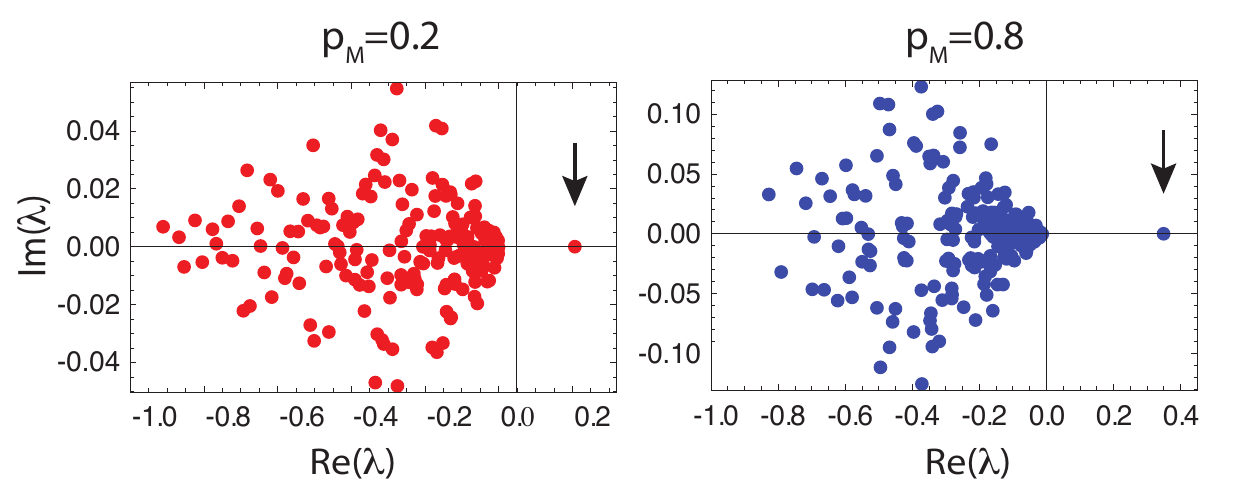}
  \caption{ The relative strengths of mutualistic and antagonistic interactions strongly affects the stability.
	The figures show the eigenvalues in the complex plane for randomly generated matrices with $S=200$ and $C=0.8$ and two values of $p_M$ (while $z_i$, $n^*_i$, $A_{ij}$, $g_{ij}$ and $e_{ij}$ are drawn from an uniform distribution between $0$ and $1$). 
	As shown in~\cite{Allesina2012}, there is always an eigenvalue which depends only on the average interaction strength, in this
	case the eigenvalue is equal to $(f_M-f_A-1)/4$ (black arrows). 
	If $f_M>f_A+1$ the system is, for a sufficiently large $S$, always unstable independently of the fraction of mutualistic interactions $p_M$ and the number of species $S$. In this case $f_M=3$ and $f_A=1$. If $f_M<f_A+1$ the destabilizing effect of $p_M$ is also observed (see Figure 4).}
\label{fig:fM-fA}
\end{figure*}

Applying this approach to the community dynamics given by Eqs.~\ref{eq:LV-def}, (\ref{eq:mat-def_mut}) and~(\ref{eq:mat-def_ant}) it turns out that $M_{ij}=-n_i^*a_{ij}$. In \cite{Mougi2012} assume that the stationary abundances $n^*$ are uniformly distributed between 0 and 1 and thus stability of an hybrid community, not only depend on the fraction of mutualistic links, but also on their strengths. In fact, in the presented community dynamics framework,  absolute strengths of the mutualistic and antagonistic interactions is defined by two parameters $f_M$ and $f_A$. As shown in~\cite{Allesina2012}, in the large $S$ limit there is always a real eigenvalue which is only determined by the average entry of the matrix
\begin{equation}
\displaystyle
\lambda_a = -\mathbb{E}(M_{ii}) + (S-1) C \mu \sim \mathbb{E}(n^*) \Bigl[ - \mathbb{E}(z) + \mathbb{E}(e)  f_M - (1-\mathbb{E}(g))  f_A \Bigr] \ ,
\label{eq:eigen-max}
\end{equation}
where  $\mu$ is the average value of the out-diagonal non-zero entries of $M$. The second equality is obtained in the large $S$ limit.
Note that the right term does not depend on $S$ and $C$, nor on $p_M$, because of the rescaling of the interactions.
By considering $z_i$, $n^*_i$, $e_{ij}$, $g_{ij}$ and $A_{ij}$ uniformly drawn between $0$ and $1$, as done in~\cite{Mougi2012}, this eigenvalue is greater than zero if $f_M > f_A+1$. 
\clearpage
Figure~\ref{fig:fM-fA} shows that the stability does depend on the choice of the relative strength of mutualistic and antagonistic interactions, even in the case of structured interactions, showing
that the prediction of equation~(\ref{eq:eigen-max}) is valid in that case too. Indeed the system is always unstable when $f_M > f_A+1$.
These results confirm the important role that mutualistic interactions play in ecological networks \cite{Bascompte2007,Melian2009} and poses the crucial problem on the estimation of these parameters from real data, a challenging but difficult task \cite{Wootton2005,Vazquez2012}. 

\subsection*{Scaling of the Interaction Strengths}

\begin{figure*}[h!]
\centering
  \includegraphics[width=0.9\textwidth]{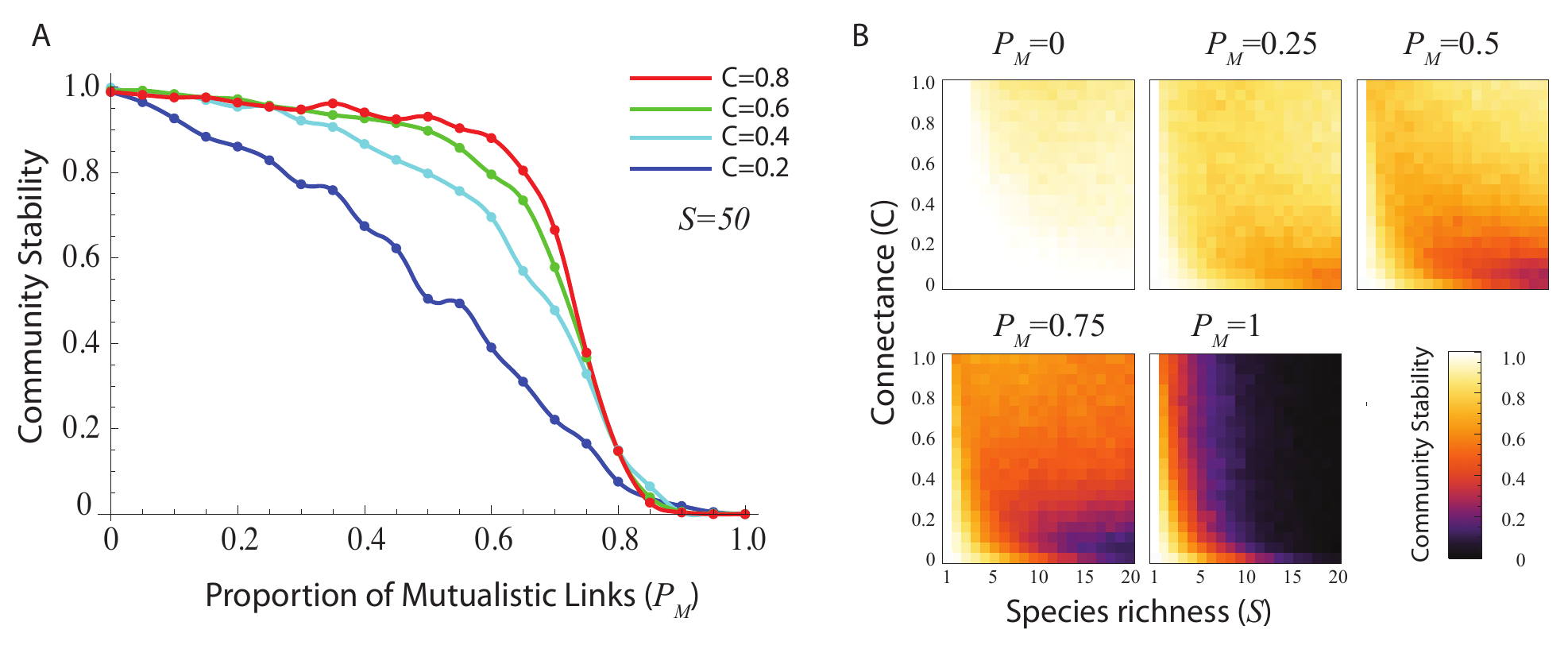}
  \caption{Relationship between complexity, stability and fraction of mutualistic links
	($p_M$) for cascade networks\cite{Allesina2012} with Holling type I linear response. The interaction
	strengths are rescaled by summing over both mutualistic partners and preys (see Eq.~\ref{eq:just-ass}, assumption $(i)$).
	Panel A shows the community stability (measured as the probability that the linearized matrix
	is stable) vs. the fraction of mutualistic links $p_M$.
	Colors indicate different values of connectance. $S$ is fixed to $50$.
	Panel B shows the complexity-stability relationship with varying $p_M$. Both the panels shows
	that mixing of interaction type in the community dynamics model given by Eq. (\ref{eq:just-ass}) and based on assumption $(i)$ does not increase stability.
	The parameters $z_i$, $n^*_i$, $A_{ij}$, $g_{ij}$ and $e_{ij}$ are drawn from an uniform distribution between $0$ and $1$, while $f_M=f_A=1$.}
\label{fig:pM}
\end{figure*}

Figure~\ref{fig:pM} shows that if a constant interacting effort is considered (assumption $(i)$) the increased stability due to mixed interactions does not hold.
This proves that the results obtained by Mougi and Kondoh are strongly affected by their peculiar assumption $(ii)$. In other words, assuming that the total mutualistic interaction strength does not depend on the relative abundance of mutualistic links (assumption $(ii)$) is the cause of the singular effect on the community stability observed in ~\cite{Mougi2012}.

\begin{figure*}[h!]
\centering
  \includegraphics[width=0.9\textwidth]{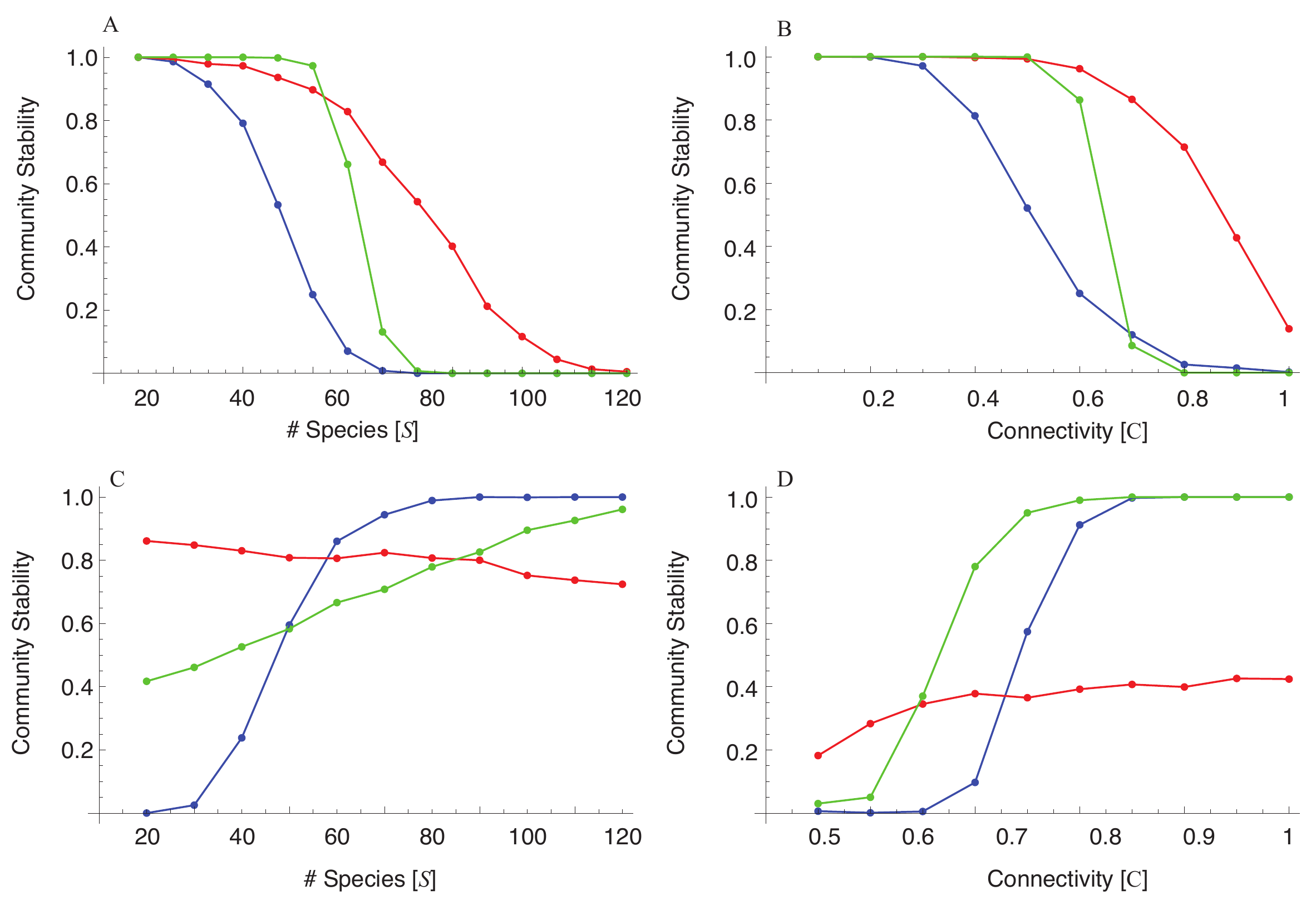}
  \caption{The rescaling of interaction strengths, and not a mixture of interaction types, produce
	a positive complexity-stability relationship.
	The stability is measured as the probability to have a positive eigenvalues. The blue curves
	are obtained with a random networks, the red curves with cascade predator-prey interactions,
	while the green ones with bipartite mutualistic matrices.
	Panel A and B show the dependence of the stability on the number of species $S$ and the connectance $C$, when the
	interactions are not rescaled. As expected the stability decreases at increased complexity.
	Panel C and D are obtained with rescaled interaction strengths as in Eqs.~(\ref{eq:mat-def_mut}) and~(\ref{eq:mat-def_ant}). The rescaling strongly affects the
	stability-complexity relation. The result obtained in~\cite{Mougi2012} is therefore not a consequence of
	mixing of interaction types (the green and red curves are obtained with purely mutualistic and cascade predator-prey interactions respectively).
	Instead it is due to the rescaling of interactions, indeed a positive complexity-stability relation
	is observable once the interactions are rescaled.
	}
\label{fig:comp-stab-rel}
\end{figure*}

Finally, a positive relation between stability and complexity is also observed in ~\cite{Mougi2012}. Again, this is a consequence of the rescaling in the interaction strength and not of the mixing of interaction types. In particular given assumption $(ii)$ it can be shown (Eq. 10 of SI in \cite{Mougi2012}) that a system with mixed interaction type is stable if a condition of the kind
\begin{equation}\label{inv_May}
\displaystyle
\frac{h(p_M)}{\sqrt{C S}} < 1
 \ ,
\end{equation}
is valid, where $h(p_M)$ is a suitable continuous function. This condition is exactly the opposite of the well known May's criterion for stability~\cite{May1972} given by Eq. (\ref{may_rel}). It is the key responsible of the Mougi and Kondoh conclusions, i.e.for hybrid ecosystem increasing $C$ and/or $S$ stabilizes the system. It is very important to note that the relation given by Eq. (\ref{inv_May}) does not arise as a consequence of the mixture of interaction types, but it is instead derived from the rescaling of interactions effort at increasing resource species. Indeed, if the interactions are rescaled following assumption $(ii)$, then the strength of of the interaction matrix elements is a decreasing function of the number of species $S$ and connectance $C$. Figure~\ref{fig:comp-stab-rel} shows that also for non-hybrid ecological networks this stabilizing effect for increasing complexity is observed.
It is exactly the imposed inverse correlation between $S C$ and the interaction strengths which produces the stabilization effect for increasing ecosystem complexity, not a mixture of interaction types. 
\subsection*{Conclusions}
 In this work we have shown that for an hybrid ecological community where a mixing of mutualistic and predator-prey interaction types are present, the complexity stability paradox  is still an issue. In particular, mutualistic interactions destabilize the system as observed by analyzing stability profiles of structured matrices \cite{Allesina2012}. The positive relation between stability and complexity observed in \cite{Mougi2012} is a consequence of a particular rescaling in the interaction strengths: by imposing that $\sigma$ scales inversely as the number of resource partners, then the stability threshold on the linearized dynamics shifts to greater values for greater complexity. However analysis of real mutualistic ecological networks support the opposite behavior: the average strength increases for increasing mutualistic partners. These results call for different principles beyond network structures and mixing of interaction types in order to understand the complexity and stability relationship in real ecological systems.
 
\section*{Acknowledgments}
AM, SS and JG acknowledges Cariparo foundation for financial support.
\bibliography{biblio}



\end{document}